\documentclass[11pt]{article}

\usepackage[preprint,hyperref]{acl}
\usepackage{times}
\usepackage{latexsym}
\usepackage[T1]{fontenc}
\usepackage[utf8]{inputenc}
\usepackage{microtype}
\usepackage{booktabs}
\usepackage{array}
\usepackage{enumitem}
\usepackage{listings}
\usepackage{xcolor}
\usepackage{url}

\definecolor{codegray}{gray}{0.95}
\title{Knowledge-as-Skill: A Structural Design for Autonomous Knowledge-Base Use by LLM Agents}

\author{Jiangxu Wu \\
  \texttt{wujx27@mail2.sysu.edu.cn}}

\newcommand{\projecturl}{\url{https://github.com/wjx-git/knowledge-as-skill}}

\begin{document}
\maketitle
\footnotetext{Project repository: \projecturl}

\begin{abstract}
Retrieval-augmented generation (RAG) is the dominant approach for giving large language models (LLMs) access to external knowledge. Its conventional ``retrieve--concatenate--generate'' pipeline, however, makes the retrieval decision on behalf of the model: evidence is retrieved and injected whether or not the question requires it. As tool use and agent loops become more reliable, decisions about whether to retrieve, what to inspect, and when to stop can be delegated to the model. This shift exposes a new bottleneck: the agent does not know what the knowledge base contains. Traditional knowledge bases are optimized for recall by a retriever. Their documents are exposed as anonymous text chunks with little information about scope, purpose, provenance, or relations, which makes them difficult for an autonomous explorer to understand.

We propose \emph{Knowledge-as-Skill}, an organization scheme in which a knowledge base is made discoverable, navigable, and self-descriptive so that an agent can use it in a manner similar to a skill. The design has three layers: a discovery layer centered on \texttt{SKILL.md}, a navigation layer consisting of one \texttt{index.md} per directory, and a knowledge layer consisting of documents with YAML frontmatter describing their topic, type, provenance, and lifecycle. The design follows the Open Knowledge Format (OKF) and the Skill protocol without requiring changes to the agent framework. We also provide \texttt{knowledge-as-skill}, a construction pipeline that converts heterogeneous collections of PDFs, Word files, web exports, and notes into this structure.

We conduct a preliminary cross-work evaluation on the WixQA enterprise customer-support benchmark. Under our current setup, the system obtains 0.889 Factuality and 0.816 Context Recall, compared with the 0.767 and 0.708 values reported by the contemporaneous Corpus2Skill work. It obtains slightly lower Faithfulness, lower Context Precision, and more interaction turns. Because the service model, prompts, and knowledge-package construction are not controlled across the two studies, these numbers are directional evidence rather than a causal comparison.
\end{abstract}

\section{Introduction}

\subsection{The awkwardness of fixed RAG pipelines}

The knowledge stored in an LLM's parameters is fixed after training. To answer questions about a particular domain, a common solution is to build a knowledge base and retrieve relevant passages before each answer. The passages are then inserted into the prompt, and the model generates an answer conditioned on them. This architecture is generally known as retrieval-augmented generation (RAG; \citealp{lewis2020rag}).

The retrieval step in conventional RAG is workflow-driven. Before answering each question, the system retrieves evidence even when the question is unrelated to the knowledge base or can be answered from the model's parametric knowledge. The workflow decides whether to retrieve, what to retrieve, and how much to retrieve. The model is primarily a recipient of the resulting context. This design is effective for many applications, but it prevents a capable model from deciding whether it needs to consult a source at all.

\subsection{A new bottleneck after retrieval becomes agentic}

Delegating retrieval decisions to the model was previously difficult for two reasons. First, models were not reliable tool users: even when a retrieval tool and usage conditions were provided, they did not consistently invoke it. Second, the dominant engineering paradigm was a fixed workflow in which each step was planned in advance. Such a workflow could not naturally represent an uncertain trajectory such as ``inspect once, read the result, and then decide what to do next.''

Both conditions have changed. Modern models have stronger reasoning and tool-use capabilities, longer contexts, and more stable agent loops. Once retrieval authority is given to an agent, however, a previously hidden problem becomes visible: \emph{the agent does not know what is in the knowledge base}.

Traditional RAG delivers a few retrieved text chunks. The chunks are often decontextualized and lack identity. The agent may not know which document produced a passage, what problem the document addresses, or whether a more relevant document is available elsewhere. Such output may be sufficient for a fixed retrieve--concatenate--answer pipeline. For an autonomous agent, the first question is different: what are these artifacts, and are they worth reading further?

A knowledge base should therefore expose documents as knowledge objects that can explain their own purpose. An agent should see a document's topic, summary, type, and relations before deciding whether to read its full text. This is closer to how a person uses a search engine: inspect a title and summary first, then open selected results.

\subsection{Knowledge-as-Skill}

We argue that a knowledge base needs an agent-oriented redesign: it should become usable as a skill. Knowledge and skills are not identical. Knowledge describes facts, whereas a skill describes a way of doing something. Our claim is more specific: when knowledge has clear boundaries, structure, documentation, and an access protocol, an agent can understand and use it \emph{like} a skill.

The resulting knowledge base should satisfy three requirements:
\begin{enumerate}[leftmargin=*]
    \item \textbf{Discoverable}: the agent can determine whether a knowledge base is relevant to the current task and whether it should use it.
    \item \textbf{Navigable}: after entering the knowledge base, the agent can progressively narrow its search through explicit structure rather than relying on a single keyword-to-chunk interface.
    \item \textbf{Self-descriptive}: each document carries metadata about its topic, purpose, type, provenance, and freshness, allowing the agent to estimate its value before reading the full text.
\end{enumerate}

We instantiate these requirements by organizing documents and indexes according to the Open Knowledge Format (OKF) and adding a Skill-compatible \texttt{SKILL.md} at the root of the knowledge base. The design, implementation, and preliminary evaluation are the contributions of this paper.

\section{Related Work}

\subsection{Retrieval-augmented generation}

RAG combines parametric generation with non-parametric evidence retrieved from an external collection \citep{lewis2020rag}. The standard recipe indexes a corpus offline, retrieves passages online, and concatenates them with the question. This architecture has been improved through query rewriting, reranking, hybrid retrieval, and iterative retrieval. These techniques operate inside the retrieval pipeline. The question considered here is orthogonal: when the model can decide what to inspect, what representation of the knowledge base supports that decision?

\subsection{Agentic retrieval}

Agentic RAG exposes retrieval as a tool and lets the model decide whether and how often to call it. This direction shares our motivation, but toolization alone does not make the backend understandable. A conventional search tool still returns decontextualized chunks. The agent remains limited to a keyword-in, passage-out interface, cannot browse the corpus structure, and cannot easily distinguish ``the knowledge is absent'' from ``the query was poorly formulated.'' Knowledge-as-Skill changes the organization of the backend so that exploration becomes a first-class access mode. This perspective is related to active retrieval, in which the generator decides when to retrieve during generation \citep{jiang2023active}.

\subsection{Agent Skills and progressive disclosure}

The Skill protocol provides a lightweight way to package capabilities. A skill is a directory containing a \texttt{SKILL.md}; its YAML frontmatter supplies discovery metadata, while the body and additional files are loaded only when needed. This is a form of progressive disclosure: information enters the model context in stages rather than all at once \citep{anthropic2025skills}. The protocol was designed primarily for procedural knowledge such as workflows and tool instructions. We apply the same mechanism to factual knowledge and argue that a useful knowledge skill needs not only a discovery file but also navigable indexes and document-level metadata.

\subsection{Corpus2Skill}

After our design and open-source implementation were completed, we became aware of the contemporaneous Corpus2Skill work \citep{sun2026corpus2skill}, which independently develops the same central idea: compile a corpus into an agent-navigable skill tree instead of using a fixed RAG pipeline. The two systems differ in their construction strategies and maintenance assumptions.

\begin{table}[t]
\centering
\small
\setlength{\tabcolsep}{3.5pt}
\begin{tabular}{p{0.22\linewidth}p{0.34\linewidth}p{0.34\linewidth}}
\toprule
Dimension & Corpus2Skill & Knowledge-as-Skill \\
\midrule
Construction & Bottom-up LLM summaries, embeddings, and iterative clustering & Specification-driven, curatorial organization under OKF \\
Navigation & Cluster-generated hierarchical indexes with exemplars & Maintained indexes containing a link and one-sentence description \\
Auxiliary mechanisms & Verification, repartitioning, soft assignment, entity index & Document frontmatter, provenance, lifecycle fields, and change log \\
Provenance & Document IDs mapped to paths & Sources recorded directly in frontmatter \\
\bottomrule
\end{tabular}
\caption{Conceptual comparison of the two independently developed approaches.}
\label{tab:comparison}
\end{table}

Corpus2Skill emphasizes automation and cold-start scalability for unstructured corpora. Our approach emphasizes controllability, auditability, and incremental maintenance for domain-bounded enterprise collections. A document can be linked from multiple indexes in our design, so cross-topic discoverability is intrinsic rather than added as a separate mechanism. The comparison is complementary rather than a claim that one construction strategy dominates the other.

\section{The Knowledge-as-Skill Design}

\subsection{Design principles}

The central principle is that knowledge should not be injected into context wholesale at the beginning of a task. It should be pulled by the agent as reasoning proceeds. Three principles follow.

\paragraph{Layered exposure.} Different files expose different information granularities. The discovery layer exposes a short trigger description, the navigation layer exposes directories and one-sentence summaries, and the knowledge layer carries complete content. Each deeper read is conditioned on the agent's preceding decision, so context cost is more closely tied to relevance.

\paragraph{Self-description.} Every document describes what it covers, which questions it can answer, which concepts it relates to, where it came from, and when it may become stale. The agent can estimate relevance before paying the cost of reading the body.

\paragraph{Separated entry and structure.} \texttt{SKILL.md} describes scope, usage, and boundaries but does not contain domain knowledge. Indexes maintain the knowledge structure independently. Content can therefore grow without destabilizing the discovery entry point.

\subsection{Discovery layer: \texttt{SKILL.md}}

The discovery layer answers: ``How does the agent know that this knowledge base exists, and when should it use it?'' The root file contains frontmatter describing the scope and trigger conditions:

\begin{lstlisting}[language={},caption={Example discovery metadata.},label={lst:skill}]
---
name: team-knowledge
description: Product decisions, engineering standards, incident response,
  and metric definitions. Use for questions about team decisions,
  release procedures, production incidents, or business metrics.
type: Knowledge Base Guide
---
\end{lstlisting}

When the knowledge base is placed in an agent framework's skills directory, its name and description can remain in the system prompt as lightweight discovery metadata. If the agent judges the task relevant, it reads the body of \texttt{SKILL.md}. The body should remain short and specify scope, exclusions, navigation rules, common entry points, freshness conventions, and stopping conditions. Explicit stopping conditions help prevent repeated exploration of an irrelevant knowledge base.

\subsection{Navigation layer: indexes and bundles}

The navigation layer answers: ``How does the agent locate knowledge after entering the base?'' A knowledge base is organized into nested directories, or knowledge bundles. Each bundle contains an \texttt{index.md} that acts as a map of that level:

\begin{lstlisting}[language={},caption={Illustrative knowledge-skill directory.},label={lst:tree}]
team-knowledge/
|-- SKILL.md
|-- index.md
|-- product/
|   |-- index.md
|   |-- pricing.md
|   `-- roadmap.md
`-- glossary/
    |-- index.md
    `-- key-metrics.md
\end{lstlisting}

Each index entry should be a link followed by one sentence of description, rather than a bare filename list:

\begin{lstlisting}[language={},caption={Illustrative navigation index.},label={lst:index}]
# Engineering

* [Incident response](incident-response.md) -- Severity,
  containment, notification, and postmortem procedures.
* [Release standards](release.md) -- Requirements for review,
  canary rollout, and rollback.
* [Service-level objectives](slo.md) -- SLI, SLO, and alert-threshold
  definitions for core services.
\end{lstlisting}

This small requirement is central to navigation. At each step, the agent chooses an exploration path from a semantic cue. Missing or vague descriptions turn navigation into random file opening. Indexes also provide an access path that complements keyword and vector retrieval. Its efficiency, especially for multi-hop questions, remains an empirical question.

\subsection{Knowledge layer: documents with frontmatter}

The knowledge layer answers: ``How can the agent estimate a document's value and provenance before reading it?'' Each document begins with YAML frontmatter. The required fields are \texttt{type}, \texttt{title}, and \texttt{description}; additional fields capture tags, sources, and lifecycle information.

\begin{lstlisting}[language={},caption={Illustrative knowledge-document metadata.},label={lst:frontmatter}]
---
type: Runbook
title: Production incident response
description: Procedures for triage, containment, notification,
  and postmortem after a production incident.
tags: [on-call, incident, production]
sources:
  - title: incident-response.docx (original)
generated:
  by: agent:mdconvert
  at: 2026-08-14T00:00:00Z
verified:
  by: human:platform-team
  at: 2026-08-15T00:00:00Z
stale_after: 2027-02-14T00:00:00Z
---
\end{lstlisting}

The description supports relevance judgments; type and tags support filtering; sources support traceability; and the generation, verification, and expiration fields support freshness-sensitive answers. A root-level \texttt{log.md} can record updates. Thus, a document carries not only content but also information about its own quality and lifecycle.

\subsection{A typical end-to-end interaction}

A representative interaction proceeds as follows:
\begin{enumerate}[leftmargin=*]
    \item The agent sees the names and descriptions of available skills and judges whether the question matches this knowledge base.
    \item It reads \texttt{SKILL.md} to learn the scope, navigation rules, and boundaries.
    \item It starts from the root \texttt{index.md} and chooses a promising bundle.
    \item It uses document frontmatter to select documents before reading their full text.
    \item If evidence is insufficient, it follows related index entries, or stops when the question is outside the knowledge-base scope.
    \item It composes an answer from the documents it actually read, consulting lifecycle metadata for time-sensitive claims.
\end{enumerate}

Every read in this process is an agent decision conditioned on prior context. The knowledge base changes from a corpus passively sliced and injected by a system into an information space that an agent can browse, assess, and revisit.

\begin{table}[t]
\centering
\small
\begin{tabular}{p{0.28\linewidth}p{0.50\linewidth}}
\toprule
Artifact & Role and reading time \\
\midrule
\texttt{SKILL.md} frontmatter & Trigger description; always available \\
\texttt{SKILL.md} body & Scope, navigation, and stopping conditions; after a match \\
\texttt{index.md} & Candidate entries and summaries; on entering a bundle \\
Document frontmatter & Topic, type, provenance, and freshness; before full-text reading \\
Document body & Complete evidence; after relevance is established \\
\texttt{log.md} & Change history; for freshness-sensitive questions \\
\bottomrule
\end{tabular}
\caption{Roles of the three layers and their typical reading times.}
\label{tab:layers}
\end{table}

\section{Construction Pipeline}

Real knowledge bases are rarely clean Markdown collections. They contain PDFs, Word files, slide decks, web exports, and meeting notes, often without a consistent structure. We implemented \texttt{knowledge-as-skill}, a construction skill that converts such collections into OKF- and Skill-compatible knowledge packages. OKF makes the knowledge readable, linked, and traceable; the Skill protocol makes the package discoverable and loadable at the appropriate time.

The pipeline has five stages. First, it enumerates the source files, keeps existing Markdown, and converts other formats to Markdown with \texttt{mdconvert}, a wrapper around MarkItDown, while preserving the directory structure. Second, it generates frontmatter. The required fields are especially important because the agent sees them before the body. The pipeline also records original sources and lifecycle fields. Third, it generates an \texttt{index.md} for each bundle. The initial index is drafted from document metadata and can be revised by a human. Fourth, it generates the root \texttt{SKILL.md}, including scope, trigger description, navigation rules, common entry points, and freshness guidance. Finally, it validates the result with \texttt{okfcheck}. Standard validation checks OKF consistency; \texttt{--agent-ready} additionally checks for a discovery file, required metadata, and missing indexes.

After construction, the package can be placed in the agent framework's skills directory. Incremental maintenance is similarly layered: a new document needs frontmatter and one index entry, while \texttt{log.md} records the change.

\section{Preliminary WixQA Evaluation}

\subsection{Research questions}

We ask three questions:
\begin{itemize}[leftmargin=*]
    \item \textbf{RQ1 (feasibility):} Can a knowledge skill support end-to-end knowledge access for enterprise question answering?
    \item \textbf{RQ2 (behavioral cost):} What trade-off does hierarchical navigation create among evidence coverage, interaction length, and context focus?
    \item \textbf{RQ3 (cross-work reference):} How do our results differ in direction from the values reported by Corpus2Skill on the same benchmark?
\end{itemize}

\subsection{Dataset and system}

We use WixQA, an enterprise customer-support benchmark built from 6,221 Wix Help Center articles. We evaluate the 200-question expert-written subset, where each example contains a question, a reference answer, and one to three gold documents. The dataset choice follows Corpus2Skill, enabling a directional reference to that work's reported values \citep{sun2026corpus2skill}.

The evaluated system is a minimal agent loop. Its system prompt contains the full \texttt{SKILL.md} and a task instruction to answer only from the knowledge base, use documents it actually read, and submit the answer when finished. We intentionally avoid embedding navigation strategies or answer-format constraints so that navigation behavior primarily comes from the knowledge structure. The agent has three read-only tools: \texttt{list\_dir}, \texttt{read\_file}, and \texttt{submit\_answer}. The maximum trajectory length is 15 turns. The answering model is DeepSeek-Flash and the judge model is K3.

\subsection{Metrics and protocol}

To reduce implementation differences, we reuse the metric implementation and evaluation prompts released with Corpus2Skill. This does not make the comparison controlled: models, prompts, and package-construction procedures differ. Token F1 is computed by a SQuAD-style token-level comparison. Factuality, grounded Faithfulness, Context Recall, and Context Precision are scored by an LLM judge on a five-point scale and normalized. Hallucination Rate is the fraction of questions whose raw Faithfulness score is at most three. Turns are measured from agent trajectories.

\begin{table}[t]
\centering
\small
\setlength{\tabcolsep}{4pt}
\begin{tabular}{lrr}
\toprule
Metric & Ours & Corpus2Skill \\
\midrule
Token F1 & 0.412 & 0.456 \\
Factuality & \textbf{0.889} & 0.767 \\
Faithfulness (grounded) & 0.810 & 0.859 \\
Context Recall & \textbf{0.816} & 0.708 \\
Context Precision & 0.761 & 0.829 \\
Hallucination Rate & 4.5\% & 4.5\% \\
Turns (mean) & 4.015 & 2.38 \\
\bottomrule
\end{tabular}
\caption{Preliminary WixQA results. Corpus2Skill values are reported values, not measurements reproduced under our environment.}
\label{tab:results}
\end{table}

\subsection{Results and analysis}

Table~\ref{tab:results} shows a mixed benefit--cost pattern. Factuality is 0.889 versus the reported 0.767 (+12.2 percentage points), and Context Recall is 0.816 versus 0.708 (+10.8 points). Under our conditions, the agent read documents covering more reference-answer content and produced answers judged more factually correct. These results are consistent with the hypothesis that explicit navigation can expand evidence discovery, but they do not establish that the structure alone caused the improvements.

Grounded Faithfulness is lower (0.810 versus 0.859), while Hallucination Rate is identical at 4.5\%. This combination illustrates that Factuality and Faithfulness measure different properties. An answer may agree with the reference answer while making weaker use of, or weaker textual connections to, the evidence actually read. Both metrics are LLM-judge scores, so small differences should be interpreted cautiously.

Context Precision is lower (0.761 versus 0.829), and the agent uses more turns on average (4.015 versus 2.38). These outcomes are consistent with the cost of progressive exploration: the agent reads more intermediate or ``passing'' documents while narrowing the search. Token F1 is also lower (0.412 versus 0.456), but this lexical metric is sensitive to wording, answer length, and formatting. Our minimal prompt imposes no answer-length or formatting constraint, and the service models differ.

Overall, the most important result is the combination of higher Factuality and Recall with lower Faithfulness and Precision and longer trajectories. It should be treated as a hypothesis-generating cross-work observation, not evidence that curatorial organization is causally superior. A controlled study should hold the model, prompts, source corpus, document conversion, and evaluation implementation constant.

\section{Limitations and Future Work}

This paper presents a structural design and a preliminary evaluation rather than a controlled benchmark study. The comparison with Corpus2Skill is limited by different service models, prompts, construction procedures, and potentially different runtime details. The current experiment also does not compare Knowledge-as-Skill against a fixed-RAG baseline under the same agent, nor does it isolate the contribution of discovery metadata, indexes, frontmatter, and lifecycle fields through ablations. LLM-judge metrics may be sensitive to prompt and model choice.

The design introduces its own engineering costs. Curated indexes require maintenance, descriptions may be incomplete or biased, and additional navigation turns consume latency and context. Automatic construction can reduce this burden but may produce inaccurate summaries or metadata. Future work should conduct controlled component ablations, measure latency and token cost, test changes over time, evaluate cross-index navigation on multi-hop questions, and study hybrid systems that combine structured navigation with conventional retrieval.

\section{Conclusion}

We introduced Knowledge-as-Skill, a structural design for allowing LLM agents to discover, navigate, and use external knowledge without having a system decide in advance which passages to inject. The design separates discovery through \texttt{SKILL.md}, navigation through hierarchical \texttt{index.md} files, and evidence through self-describing documents with frontmatter. We also described an automated construction pipeline for converting heterogeneous knowledge collections into this structure.

On the WixQA benchmark, our preliminary setup obtains 0.889 Factuality, 0.816 Context Recall, and a 4.5\% Hallucination Rate, at the cost of 4.015 average interaction turns. Compared with values reported by Corpus2Skill, it shows higher Factuality and Context Recall but lower Faithfulness and Context Precision. These results motivate controlled experiments rather than a definitive ranking. The broader claim is architectural: when an agent is responsible for deciding what knowledge to read, the knowledge base should expose structure, identity, and provenance instead of presenting only anonymous retrievable chunks.

\bibliography{references}

\end{document}